Title: **Detection of acoustic events in Lavender for measuring the xylem vulnerability to embolism and cellular damages.**

Corresponding author: Guillaume Charrier

Université Clermont Auvergne, INRAE, PIAF, F-63000 Clermont–Ferrand, France. guillaume.charrier@inrae.fr

Authors names and affiliations: Lia Lamacque[1,2], Florian Sabin[1], Thierry Améglio[1], Stéphane Herbette[1], Guillaume Charrier[1*]

[1] Université Clermont Auvergne, INRAE, PIAF, F-63000 Clermont–Ferrand, France.

[2] Institut Technique Interprofessionnel Plantes à Parfum, Médicinal, Aromatiques et Industrielles, 26740 Montboucher-sur-Jabron

**Short title:** Acoustic detection of drought-induced damages

**One-sentence summary:** The discrimination between two independent phases of acoustic emissions reveals the succession of hydraulic failure and cellular damages leading to drought-induced plant mortality.

**Author contributions:** G.C., F.S., L.L., S.H. and T.A. conceived the experiment; L.L. and F.S. collected the data; L.L., F.S., and G.C. analyzed the data; L.L. and G.C. wrote the article with help from S.H. and T.A.; all authors contributed critically to the drafts and gave final approval for publication. G.C. agrees to serve as the author responsible for contact and ensures communication.

**Funding information:** This work was supported by the French Ministry of Agriculture (CASDAR grant no. 5638; "RECITAL"), the Association Nationale de la Recherche et Technologie (CIFRE grant no. 2016/1280) and the French Research Institute for Perfurme, Medicinal and Aromatic Plants (Iteipmai).

**Aknowledgments:** The authors thank Johanna Roustant (EARL Les roustanets, Les Granges-Gontardes) for providing plant materials; Patrice Chaleil and Aline Faure (Université Clermont Auvergne, Institut National de Recherche pour l'Agriculture, l'Alimentation et l'Environnement, Unité Mixte de Recherche PIAF) for plant care during the growing period; Pierre Conchon for his help in micro-CT measurements.

**Abstract (290 words)**

Acoustic emission analysis is a promising technique to investigate the physiological events leading to drought-induced injuries and mortality, as the measurements are real time and non-invasive. However, the nature and the source of the acoustic emissions are not fully understood and make the use of this technique difficult as a direct measure of the loss of xylem hydraulic conductance under drought stress. In this study, acoustic emissions were recorded during severe dehydration in lavender plants (*Lavandula angustiolia* Mill) and compared to the dynamics of embolism development and cell lysis. The timing and characteristics of acoustic signals from two independent recording systems were compared by principal component analysis. In parallel, changes in water potential, branch diameter, loss of hydraulic conductance and electrolyte leakage were measured to quantify the response to drought and related damages. Two distinct phases of acoustic emissions were observed during dehydration. The first phase was associated with a rapid loss of diameter and a significant increase in loss of xylem conductance (90%). The second phase was mostly associated with a significant increase in electrolyte leakage whereas diameter changes were slower. This phase corresponds to a complete loss of recovery capacity. The acoustic signals of both phases were discriminated by the third and fourth principal components. The loss of hydraulic conductance during the first acoustic phase suggests the hydraulic origin of these signals (*i.e.* cavitation events). For the second phase, the signals showed much higher variability between plants and acoustic systems suggesting that the sources of these signals may be plural, although likely including cellular damage. A simple algorithm was developed to discriminate hydraulic-related acoustic signals from other sources, allowing the reconstruction of dynamic hydraulic vulnerability curves. However, hydraulic failure precedes cellular damage and lack of whole plant recovery is associated to these latter.

## Introduction

Drought stress in plants leads to a cascade of physiological events as water content progressively decreases. Under mild stress, stomata close, thus limiting water losses through decreased transpiration. However, bulk water content continuously decreases *via* cuticular transpiration (Martin-StPaul et al., 2017). Under extreme drought conditions, the critical branch diameter at which the plant loses its ability to rehydrate is an indicator of plant mortality (ca. 20% in Lavander; Lamacque et al., 2020). Plant mortality is observed after the complete xylem hydraulic failure, when cellular damages increases dramatically (ca. 75% Lamacque et al., 2020).

In the hydraulic system, water flows under a metastable state, according to the gradient in water potential (Dixon and Joly, 1895). When water losses (evapotranspiration) exceed water supply (root water uptake) and intrinsic pools (capacitance), the tension in the xylem sap increases. Above a critical value, xylem sap metastability is broken resulting in a sudden formation of gas bubbles (i.e. cavitation) that expand and embolize the xylem conducting elements (Lewis, 1988). Xylem embolism therefore decreases the sap flow thereby inducing the dehydration of distal organs (e.g. leaves and buds; Tyree and Sperry, 1989). Xylem embolism is thus tightly correlated to plant mortality (Brodribb and Cochard, 2009; Barigah et al., 2013; Anderegg et al., 2016). Although the water potential inducing more than 88% loss of hydraulic conductivity has long been considered as an important trigger of plant mortality in angiosperm under drought stress (Barigah et al., 2013; Urli et al., 2013; Li et al., 2016), recent studies highlighted more equivocal results and thus the need to further investigate the mechanisms of drought-induced mortality (Nardini et al., 2013; Hammond et al., 2019).

Plant mortality is indirectly assessed through the inability to regrow or resprout once the stress is released, until the following spring. The viability of meristematic cells is thus key to predict plant survival (Guadagno et al., 2017). However, hydraulic failure and cell viability have been rarely studied together (Ganthaler and Mayr, 2015) although their interaction is probably key to predict drought mortality mechanisms (Charrier et al., 2020; Lamacque et al., 2020).

Numerous methods have been used to quantify hydraulic failure such as the gravimetric method (Sperry et al., 1988) or the pressure sleeve (Ennajeh et al., 2011). However, these methods are destructive and may produce unrealistic values by the induction of bubbles at sample ends whenever the tension of the sap is not released by successive cuttings (Wheeler et al., 2013). The X-ray microtomography allows spatialized observation of embolized xylem on a cross section (Cochard et al., 2015; Choat et al., 2016; Nolf et al., 2017). However, the exposure to high energy radiation has deleterious consequences on cellular viability (Petruzzellis et al., 2018). Non-invasive methods have been more recently developed to measure the changes in light transmission caused by air spreading within xylem tissue of leaves and stems (Brodribb et al., 2016; Brodribb et al., 2017).

Cellular damages can be assessed through the amount of electrolytes released in solution or by staining techniques to detect living cells and therefore dead cells (Lamacque et al., 2020). This technique has been used to study the effect of various stress factors such as chilling (Herbette et al., 2005; Mai et al., 2009), frost (Charrier and Améglio, 2011; Guàrdia et al., 2016) or drought stress (Guadagno et al., 2017; Lamacque et al., 2020). The variation of branch diameter allows a quantitative assessment of damages and has been used to study the effect of frost (Améglio et al., 2003) or drought stress (Lamacque et al., 2020). The amount of damages can indeed be evaluated from the ratio of initial diameter to recovered diameter after stress has been released (Lamacque et al., 2020).

The detection of acoustic events have been used as a non-invasive technique to analyze plant's response to drought (Milburn and Johnson, 1966; Tyree and Dixon, 1983). Acoustic events have been recorded under drought and frost stress, on potted and naturally growing plants (Lo Gullo and Salleo, 1992; Rosner et al., 2006; Vergeynst et al., 2014; Charrier et al., 2017). The nucleation of bubbles in the xylem elements, observed by microscopic observations, generates acoustic events (Ponomarenko et al., 2014). However, the cumulated number of acoustic events usually exceeds the theoretical number of xylem conducting elements in the sample (Rosner et al., 2006; Kasuga et al., 2015; Vergeynst et al., 2015). Furthermore, acoustic events are also recorded after xylem is fully embolized (Wolkerstorfer et al., 2012; Nolf et al., 2015). The cumulative rate of acoustic events is therefore not always proportional to the loss of hydraulic conductivity as cavitation events in non-conductive cells do not have any influence on xylem hydraulic (Nolf et al., 2015). The use of this technique has thus been limited, as a complete overview of acoustic sources is not clearly identified (Wolkerstorfer et al., 2012; Kasuga et al., 2015).

Taking acoustic events characteristics into account, such as amplitude, energy and frequency, can be a promising way to improve the accuracy of the information given by this technique and thus strenghen its predictive ability. Energy-wheighed acoustic events seem more relevant from an hydraulic point of view (Mayr and Rosner, 2011; Kasuga et al., 2015). During the freezing of walnut stems, Kasuga et al. (2015) identified that two types of acoustic events were generated by frost-induced embolism and by intracellular freezing. In dehydrating branches from grapevine, three clusters of events have been characterized: high, middle and low-frequency events (Vergeynst et al., 2016). High frequency signals were assumed to be generated by capillary action of water and fast contraction of the bark, while low frequency signals were generated by micro-fractures (Vergeynst et al., 2016). Mid-frequency signals seem linked to hydraulic failure and likely generated by cavitation events.

To monitor drought-induced damages (i.e. hydraulic and cellular damages), we hypothesized that acoustic emissions would be generated at distinct moments, in relation with hydraulic failure then cellular damages as observed in Lavender (Lamacque et al., 2020). We expect that a first phase of acoustic event would be related to loss of hydraulic conductance and a second phase to cellular damages. Multivariate signal analysis would allow defining clusters of acoustic emissions related to hydraulic failure and cellular damages. During an extreme dry-down, continuous monitoring of acoustic emissions and stem diameter were related to relevant physiological parameters: water potential, electrolyte leakage and loss of hydraulic conductivity. Based on intrinsic acoustic characteristics, the multivariate signal analysis of acoustic events would allow discretizing acoustic events from hydraulic and non hydraulic origin and thus simulate the dynamic of both the loss of hydraulic conductance and the cellular damages during a severe dehydration.

**Results**

During dehydration, lavender plants exhibited two distinct phases of high acoustic activity (> 1 AE.min$^{-1}$) clearly separated in time by a relatively less active phase (Fig. 1). The first phase occurred during a strong decrease in diameter, whereas during the second phase, change in diameter decrease was much lower, reaching almost steady values. The highest activities during the two phases ($AE_1$ and $AE_2$) occurred at relatively similar percent loss in diameter (PLD) across plants: 11.75 ± 0.84 and 22.26 ± 1.10% PLD for $AE_1$ and $AE_2$, respectively.

During the two periods of intense acoustic activity, loss of hydraulic conductivity (PLC) and electrolyte leakages (EL) increased independently (Fig. 2A and B). After the first acoustic phase, PLC reached 87.32 ± 3.61% and EL 47.22 ± 10.03%. After the second acoustic phase, PLC remained high (97.37 ± 1.14%), while EL reached 75.37 ± 9.31%. Depending on PLD, PLC and EL both fit a sigmoidal function (pseudo-$r^2$ = 0.85 and 0.75 for PLC and EL, respectively (Lamacque et al., 2020)). The water potential decreased with increasing PLD according to an exponential model (pseudo-$r^2$ = 0.46, Fig. 2B), reaching lower values than -9MPa (*i.e.* minimum measurable water potential) after $AE_2$.

The contrasted dynamics in PLC and EL suggest that two distinct acoustic phases may result from different phenomenon. The variability across extracted characteristics from each acoustic events recorded using a PCI2 recording system was explored through multivariate principal component analysis (PCA) with respect to acoustic phase (Fig.3A and Supplemental Fig.S1). The dataset exhibits a Kaisier-Mayer-Olkin index of 0.652 suggesting that PCA had a good ability to summarize the information contained in the data and that PCA is thus usable on such dataset. The four first component contributed to 71% of cumulated variance (from 33 to 10% for each component; Table 1). Signal strength, energy, counts and duration of the acoustic event mainly contributed to the first component (Dimension 1: $Dim_1$; Supplemental Table S1). Frequency related parameters, but initiation frequency mainly contributed to the second component (Dimension 3: $Dim_2$; Supplemental Table S1). However, $Dim_1$ and $Dim_2$ did not highlight any difference across acoustic phases (Supplemental Fig. S1). Absolute energy and parameters related to the beginning of the acoustic event (initiation frequency, rise time and rise angle) mainly contributed to the third component (Dimension 3: $Dim_3$; Fig. 3B and Supplemental Table S1). Initiation frequency, rise angle, average frequency and counts to peak mainly contributed to the fourth component (Dimension 4: $Dim_4$). According to $Dim_3$ and $Dim_4$, acoustic phases were discriminated from top left to bottom right (along the y = - x line, Fig.3A). Signals from $AE_1$ were located all along this line, whereas signals from $AE_2$ were only located in the top left quarter (*i.e.* positive latent variable 3 $LV_3$ and negative latent variable 4 $LV_4$, with respect to $Dim_3$ and $Dim_4$, respectively; Fig. 3A).

Based on $LV_3$ and $LV_4$, an algorithm predicting the position of each acoustic event within the latent variable space was developed to discriminate acoustic events that occur only during cavitation events. Based on this decision rules, filtered acoustic emissions did exhibit a sigmoid pattern in response to loss of diameter with $PLD_{50}$ equal to 11.0 ± 0.3% (Fig. 4A). Thanks to the non-linear relation (described in Fig. 2B), the water potential inducing 50% of cavitation events was computed as $\Psi_{AE50}$ = -4.06 ± 0.13 MPa (the water potential inducing 50% cavitation events determined via acoustic emissions; Fig. 4B).

The algorithm developed from PCI2 system was applied to the signals recorded from the second ultrasound acoustic system called Samos. As observed with signals from PCI2, filtered acoustic emissions also exhibited a sigmoid pattern in response to loss of diameter with similar $PLD_{50}$ ($PLD_{50} = 11.3 \pm 0.9\%$; Figure 4C). With respect to water potential, $\Psi_{AE50}$ was also similar to the one computed with PCI2 signals: $\Psi_{AE50} = -4.33 \pm 0.39$ MPa (Fig. 4D). Finally, the vulnerability curves derived from both acoustic systems overlapped (Fig. 5) and $\Psi_{PLC50} = -2.54 \pm 0.2$ MPa (the water potential inducing 50% loss of hydraulic conductance).

Filtered out signals (*i.e.* negative $LV_3$ or positive $LV_4$), named non-hydraulic acoustic events, did not exhibit a clear monotonic trend, but rather a double S shaped function in relation to PLD or $\Psi$ (Fig.6). Furthermore, the variability across replicates was much higher than filtered signals. However, the dynamic of these signals was consistent with the increase in electrolyte leakage (Fig.6).

## Discussion

Continuous monitoring of stem diameter and acoustic events allowed distinguishing two key events of drought-induced damages. The detection of distinct phases of acoustic events provided an exceptional opportunity to discriminate between hydraulic related acoustic events and other sources to monitor drought-induced damages. A single phase of acoustic events is usually recorded during dehydration (Rosner et al., 2006; Rosner et al., 2009). However, beyond the peak in acoustic activity, numerous emissions were recorded following a long tailed distribution (Wolkerstorfer et al., 2012; Nolf et al., 2015). In the present study, the first phase of acoustic events $AE_1$ occurred during the main and quick shrinkage of the stem while the second phase occurred when diameter change became lower. The transition between the two acoustic phases corresponded to a breakpoint of diameter change at PLD 10-15% while xylem embolism was almost total (PLC > 90%). We therefore suggest that most signals corresponding to cavitation events were recorded during $AE_1$. However, cellular damages were also increasing during $AE_1$, not only cavitation events are thus likely to be recorded during this phase.

During the second phase $AE_2$, since the shrinkage goes on and the branch becomes completely dry, other events than cavitation are likely to generate acoustic events such as cracks (De Roo et al., 2016) or cell wall shrinkage (Čunderlik et al., 1996). The relation with cellular damages is not straightforward although frost-induced cellular damages has been shown to induce acoustic events (Kasuga et al., 2015). The correlation between acoustic events and cellular damages can be explained by different processes such as the membrane rupture, the intracellular cavitation or wall cracks (Sakes et al., 2016). Two successive acoustic phases were also related to stem shrinkage and to cavitation events, respectively (Vergeynst et al., 2014). However, in our study, acoustic events were recorded over a much wider range of drought stress, well beyond the threshold for hydraulic failure and cellular damages ($\Psi < -9$ MPa). As plant mortality is entailed, the membrane rupture or intracellular cavitation is likely to happen (Sakes et al., 2016). As cellular damage and mechanical constraints may be generated over the whole dehydration process and can induce acoustic events, we suggest that these signals were recorded during both acoustic phases.

The principal component analysis used on the parameters of the population of acoustic events allowed discriminating acoustic emissions related to hydraulic failure from the other sources. The first two principal components ($Dim_1$ and $Dim_2$) did not allow any discrimination across acoustic phases. We suggest that the parameters contributing significantly to these two principal components were mainly related to the structure of the plant tissue and driven by attenuation properties within plant tissue. However, the third and fourth principal components ($Dim_3$ and $Dim_4$) did allow a clear distinction between two acoustic phases $AE_1$ and $AE_2$. As all the signals corresponding to cavitation events were recorded during AE1, we suggest that the signals of positive $LV_3$ and negative $LV_4$ are generated by cavitation events.

The simple rule ($LV_3 > 0$ and $LV_4 < 0$) was used to reconstruct a vulnerability curve based on acoustic emissions. Vulnerability curves reconstructed with acoustic emissions, showed similar results between the two acoustic recording systems PCI2 and Samos, which have different frequency ranges (1kHz-3MHz and 1kHz-400kHz, respectively) with $\Psi_{AE50}$ = -4.06 ± 0.13 MPa and -4.33 ± 0.34 MPa. The diameter loss curves are also similar between the two systems, with very close $PLD_{50}$ (ca. 11%). Testing the decision rules with an independent recording system allowed validating the previously performed calibration.

The hydraulic vulnerability reported with this method is slightly lower than $\Psi_{PLC50}$ acquired using hydraulic methods (ca. -2.5 MPa; Lamacque et al., 2020). Even if each single event would represent exactly one cavitation event in a conduit (Tyree and Dixon, 1983; Tyree et al., 1984), the relation between cumulative AEs and conductivity loss will be non-linear if not all the xylem elements individually contribute equally to the total hydraulic conductivity (Cochard, 1992). The impact of an embolized vessel on the conductance would greatly vary depending on the vessel dimensions, the xylem network and the level of xylem embolism. Within the xylem, the larger vessels are the first to be embolized, while the smaller ones are the last (Lemaire et al., 2021). The first cavitation events therefore have a greater effect on conductance than the last ones, so that the loss of conductance is very high when half the vessels are embolized. This might explains why the water potential inducing 50% cavitation events ($\Psi_{AE50}$) would be lower than the one inducing 50% loss of conductance ($\Psi_{PLC50}$). Furthermore, the water potential was not measured continuously but punctually during dehydration. To represent the acoustic emissions as a function of water potential, we used the non linear relationship found between $\Psi$ and PLD. However, this translation is only an estimate and an uncertainty in the computed $\Psi$ values remains. This could contribute to the discrepancy between $\Psi_{AE50}$ and $\Psi_{PLC50}$. Similar experimentation with higher frequency in the $\Psi$ measurements could help mitigate such a difference.

Loss of hydraulic conductance during the first acoustic phase supports the hydraulic origin of these signals. For the second phase, the signals did exhibit much higher variability across plants and acoustic systems suggesting that the source of these signals may be plural, although probably including cellular damages.

The lavender used in this study were rather vulnerable to cavitation compared to other Mediterranean shrub species (Lamacque et al. 2020). Since a large plasticity was reported for this trait (Awad et al., 2010; Herbette et al., 2010), we assume that the lavender plant grown under well-watered conditions have developed more hydraulically vulnerable xylem, whereas other parameters related to cellular damages would not exhibit such a plasticity. These particular conditions probably explain why two distinct acoustic phases were observed while plants grown in the field would have exhibited overlapped AE phases (Nolf et al., 2015).

Previous results on lavender showed that the percent loss of diameter corresponding to a complete loss of rehydration capacity (PLRC) was equal to 21.3% (Lamacque et al., 2020). Here, $AE_1$ and $AE_2$ peaks were recorded on average for PLD = 11.7% and 22.3%, respectively. Based on the reported relationship between PLD and PLRC (Lamacque et al., 2020), $AE_1$ and $AE_2$ would be associated to to 30.2% and 98.1% PLRC, respectively. Hydraulic failure, determined through $AE_1$ phase is thus not a relevant signal for plant mortality even though it could lead to the plant mortality through the complete desiccation of the meristematic cells (Lamacque et al., 2020; Mantova et al., 2021). The assessment of cellular damages has been identified as relevant to predict plant mortality under extreme drought (Vilagrosa et al., 2010; Guadagno et al., 2017; Mantova et al., 2021). The detection of $AE_2$ emissions would therefore be a relevant signal to detect plant mortality.

## Conclusion

Two distinct phases of acoustic emissions were recorded during the complete dehydration of lavender. Each phase had a particular acoustic signature, highlighted by the principal component analysis method, and was related to successive drought-induced damages: xylem hydraulic failure and cellular damages. Hydraulic failure happened much earlier than cellular damage, the latter being tightly linked to a critical water content that can be reached more rapidly after xylem hydraulic failure. By combining acoustic measurements, loss of diameter, PLC and EL measurements, this study clarified the dynamics of drought-induced damages and proposed a way to reconstruct xylem vulnerability curves from the non-invasive acoustic emission analysis.

**Materials and Methods**

**Plant material**

Experiments were carried out on a clonal variety of lavender (*Lavandula angustifolia* Maillette (*L.a.*) during the summer 2018. Plants were grown from 1-year-old cuttings provided by a lavender producer in Les-granges-Gontardes (N 44° 24′ 57.24″, E 4° 45′ 47.304″, 100 m *a.s.l*). In January 2018, eight plants were potted in 10 litre pots filled with 5,700 g of soil recommended for aromatic plant cultivation (Klasmann code 693: medium fibrous structure, pH 6 ± 0.3, mainly composed of blond sphagnum peat and coconut fibers for optimal ventilation) and grown in a greenhouse for six months where they were daily watered at field capacity. All eight plants were similar with respect to their morphology (height, architecture, leaf density). Temperature and relative humidity were monitored in the greenhouse and the cooling system started when temperature reached 35°C.

**Dehydration treatment**

Eight plants were dehydrated until extreme desiccation, as described in Lamacque et al. (2020). Plants were uprooted, put in a temperature-controlled chamber at constant temperature (25°C) and light levels (two lights of 25W and 172lm) and dehydrated until the stem diameter remained constant for at least 24h and no acoustic emissions were recorded within this period. The typical duration of the experiment was approx. 10 days.

**Acoustic emission analysis**

AE was monitored with a PCI-2 system (Physical Acoustics, PAC18-bit A/D, 1kHz-3MHz) and SAMOS (1kHz-400kHz) connected to broad range acoustic sensors (150–800 kHz ISD9203B) through a preamplifier set to 40 dB. One sensor per individual was directly mounted on a branch where the bark was removed along about two cm. The debarked surface was covered with silicone grease to prevent further water loss and optimize acoustic coupling. The sensor was tightly clamped to the debarked part of the sample. The acoustic detection threshold was set to 40 dB. Acoustic coupling was tested using the Hsu-Nielsen method (lead break; Kalyanasundaram et al., 2007; Charrier et al., 2014) at a distance of 1 cm from each sensor, and sensors were reinstalled when the signal amplitude was below 75 dB. AE and analysis were performed using AEwin software (Mistras Holdings). Acoustic activity was calculated as AE.min$^{-1}$ (averaged across 6 min). Five plants were analyzed using the SAMOS system and five using the PCI2 system. For a technical repetition, two plants were connected to both Samos and PCI2 systems. The experimental device is represented in Supplemental Figure S2.

Each acoustic wave was analyzed by the AEwin software and 9 relevant characteristics were extracted from the original signal (Amplitude: Amp, Energy, Absolute energy: Abs Energy, Signal Strength: Sig Strength, Rise Angle, Duration, Rise time, Counts to peak: P Count and Counts). The acoustic wave was also transformed using fast Fourier transform to compute 6 frequency related characteristics (Average frequency: A Frq, Frequency centroid: C Frq, Initiation frequency: I Frq, Peak frequency: P Frq, Reverberation frequency: R Frq, Weighed peak frequency: WPF).

**Monitoring of branch diameter variations**

Branch diameter was continuously measured using miniature displacement sensors with a friction free core glued to the bark (Supplemental Figure S5) and LVDT (model DF2.5 and DF5.0; Solartron Metrology, Massy, France) connected to a wireless PepiPIAF system (Capt-connect, Clermont-Fd, France). Straight and unbranched sections of main branches longer than five cm were randomly chosen to mount LVDT dendrometers by a custom-made stainless Invar (alloy with minimal thermal expansion) holder adapted for lavender. At the end of the dehydration, the final branch diameter was measured using a calliper (Burg Wächter, 0.01mm accuracy) at the location of friction-free core of LVDT dendrometers measurement. One LVDT per individual lavender were mounted and branch diameter recorded at five minutes intervals until the end of the experiment.

The PLD was calculated from branch diameter variation during drought stress, according to Lamacque et al., (2019) for each plant as :

$$PLD = 100 \times \frac{D_{max} - D_{min}}{D_{max}} \qquad (1)$$

where $D_{max}$ and $D_{min}$ are the maximum (before the dehydration) and the minimum (at the end of the dehydration) branch diameter, respectively.

At the end of the dehydration, the PLD was considered maximum ($PLD_{max}$).

**Water potential**

Water potential was measured regularly over the course of the dehydration using a Scholander-type pressure chamber (PMS Instrument, Albany, OR, USA). Measurements were carried out on *ca.* 5-10 cm long upper branch segments bearing several leaves. For each kinetics point, measurements were performed on each plant three times.

**Xylem embolism**

Loss of hydraulic conductance were made before the plants were uprooted and then two or three times during dehydration. Xylem embolism on stems was measured using a xylem embolism meter (XYL'EM, Bronkhorst, Montigny-les-Cormeilles, France). The entire inflorescence stems of about 30 cm long were collected and put in wet black plastic bags and brought them immediately to the laboratory for measurements. Segments of 2 cm were cut under water and fitted to water-filled tubing. One end of the stem segment was connected to a tank of de-gassed, filtered 10mM KCl and 1mM $CaCl_2$ solution. The flux of the solution was recorded through the stem section under low pressure (60 – 90 mbar) and the initial hydraulic conductance ($K_i$) scored. Then, the stem was perfused at least twice for 10 sec then 2 min at 1 bar until the hydraulic conductance no longer increased in order to remove air from embolized vessels and to determine the maximum conductance ($K_{max}$). The percentage loss of hydraulic conductance (PLC) was determined as:

$$PLC = 1 - \frac{K_i}{K_{max}} \qquad (2)$$

PLC was measured on each individual plants with at least 3 repetitions for each individual.

In addition, the PLC of three plants was determined by using X-ray microtomography (Micro-CT), a technique that allows the evaluation of embolism formation and spreading by direct observation (Choat et al., 2016). Thus, each branch was adjusted to five cm length with a razor blade, immersed in wax to avoid dehydration during scanning and placed in an X-ray microtomograph (Nanotom 180 XS, GE, Wunstorf, Germany) at the PIAF laboratory (INRAE, Clermont-Ferrand, France). For the micro-CT image acquisition and image combination, the field of view was adjusted to 5.1 x 5.1 x 5.1 mm3 and the X-ray source set to 60 kV and 240 µA. For each ca. 21 min scan, 1,000 images were recorded during the 360° rotation of the sample. The microtomography scans were recontruscted in three-dimension (3D) with a spatial resolution of 2.5µm/voxel and one transverse 2D slice was extracted from the middle of the volume using Phoening datosx 2 software (General Electric, Boston, MA, USA). After scanning the sample, a second scan was performed after the sample was cut in the air just below the scanned area, inducing air entry in the remaining functional conduits and therefore 100% of embolized xylem in the area. PLC was then calculated by comparing the area occupied by embolized vessels measured after the first scan and the total xylem conductivity area (Ae and Ac, respectively) as:

$$PLC = \frac{Ae}{Ac} \qquad (3)$$

PLC determined by using Micro-CT was measured three times during the dehydration: before the start of dehydration, after the first peak of acoustic emissions and after the second peak of acoustic emissions, in order to validate the results obtained with the Xyl'Em.

**Electrolyte leakage**

Electrolyte leakage was measured two or three times during dehydration to evaluate cellular damages induced by various stress (see *e.g.* Herbette et al., 2005). To assess drought-induced cellular damages, the electrolyte leakage (EL) was measured. Samples were cut into 5 cm long sections and immersed into 15 ml of distilled-deionized water. Vials were shaken for 24 h at +5 °C in the dark (to limit bacterial growth) on a horizontal gravity shaker (ST5, CAT, Staufen, Germany). The electric conductivity of the solution was measured ($C_1$) at room temperature using a conductimeter (Held Meter LF340, TetraCon® 325, Weiheim, Germany). After autoclaving at +120 °C for 30 min and cooling down to room temperature, the conductivity was measured again ($C_2$). Relative EL was calculated as:

$$REL = \frac{C_1}{C_2} \times 100 \quad (6)$$

To normalize the REL, 10 control samples and 10 samples were frozen at -80°C and used to have a reference for 0 ($REL_{WW}$) and 100% cellular damages ($REL_{-80}$), respectively. An index of damages $I_{Dam}$ was computed as:

$$I_{Dam} = \frac{REL - REL_{WW}}{REL_{-80}} \quad (4)$$

**Statistical analysis**

Statistical analyses were performed using the RStudio software (under R core version 4.0.3, R Development Core Team, 2020). The nls function was used to fit the relations among PLC, EL, ψ and PLD. Dataset of acoustic emissions from the PCI2 system was subjected to principal component analysis (PCA). The Kaisier - Mayer - Olkin index (KMO) was calculated to measure the sampling adequacy to PCA (the ability to summarize the information contained in the dataset).

**Table 1.** Principal component analysis performed on acoustic characteristics of signals recorded during dry-down of Lavender.

| Dimension | Variance | Variance (%) | Cumulated variance |
|---|---|---|---|
| 1 | 4.9 | 33.3 | 33.3 |
| 2 | 2.56 | 17.1 | 50.3 |
| 3 | 1.64 | 10.9 | 61.2 |
| 4 | 1.47 | 9.8 | 71.0 |
| 5 | 1.05 | 7.0 | 78.0 |
| 6 | 0.91 | 6.1 | 84.0 |
| 7 | 0.66 | 4.4 | 88.4 |
| 8 | 0.56 | 3.8 | 92.2 |
| 9 | 0.46 | 3.1 | 95.3 |
| 10 | 0.27 | 1.8 | 97.1 |
| 11 | 0.22 | 1.4 | 98.5 |
| 12 | 0.15 | 1.0 | 99.5 |
| 13 | 0.08 | 0.5 | 100 |
| 14 | 0 | 0 | 100 |
| 15 | 0 | 0 | 100 |

**Figure caption**

**Figure 1.** Acoustic activity (acoustic event per minute; black dots) and change in branch diameter (red lines) during dehydration of eight uprooted lavender plants under constant temperature (25°C). The number and the letter represent the plant and the acoustic recording system (P: PCI2 and S: Samos), respectively. The dotted lines represent the time at highest peak acoustic activity for two phases ($AE_1$ and $AE_2$).

**Figure 2. Upper panel.** Transverse section of inflorescence stems (A-C) and branches (D-E) by high resolution computed tomography at three phases during dehydration: initial phase ($T_0$; $\Psi = -1.05 \pm 0.12$ MPa), after the first peak of acoustic emissions ($AE_1$; $\Psi = -4.4 \pm 0.03$ MPa) and after the second peak of acoustic emissions ($AE_2$; $\Psi < -9$ MPa). Dark areas represent low-density areas *i.e* embolized vessels and pith. PLC is the loss of xylem hydraulic conductivity for each phase (mean ± SE from n=3 for $T_0$ and $AE_2$, and 5 for $AE_1$). **Lower panel.** PLC, water potential (wp), percent cell lysis (EL) depending on percentage loss of diameter (PLD) during dehydration. PLD at $AE_1$ and $AE_2$ are represented by vertical dotted and black lines, respectively (SE in shaded area). See Lamacque et al. (2020) for further explanations.

**Figure 3. A.** Principal Component Analysis (PCA) based on characteristics of recorded acoustic emission during dehydration of uprooted lavender with PCI2 system. Black and red dots represent the signals recorded during distinct acoustic phase ($AE_1$ and $AE_2$, respectively). Dimensions 3 and 4 are represented as they maximize the discrimination between $AE_1$ and $AE_2$ (the other combinations are represented in the supplementary figure S1). **B.** Contribution of each acoustic characteristic to the Dim 3 and 4.

**Figure 4.** Cumulated hydraulically related acoustic events based on Dimension 3 and 4 depending on the percentage loss of diameter (PLD; A and C) and the water potential (B and D). Upper and lower panels represent the signals recorded by PCI2 and Samos recording systems, respectively.

**Figure 5.** Cumulated hydraulically related acoustic events from two independent recording systems PCI2 (black line) and Samos (red line) and loss of hydraulic conductance measured on branches (blue points and line).

**Figure 6.** Non-hydraulic acoustic events from two independent recording systems PCI2 and Samos, in black and red, respectively. Relative cellular damages measured by electrolyte leakage method during the dehydration are represented in blue.

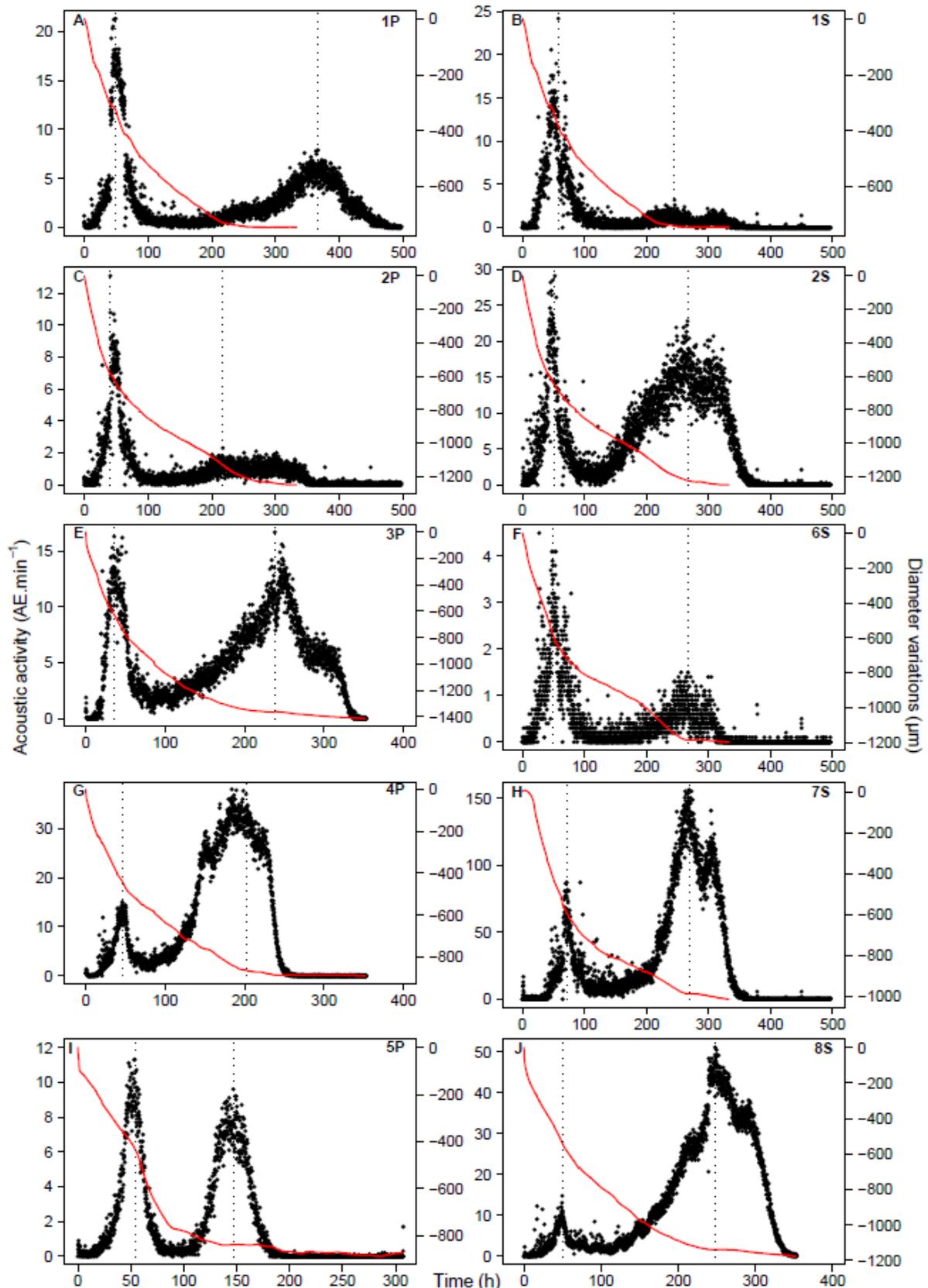

**Figure 1.** Acoustic activity (acoustic event per minute; black dots) and change in branch diameter (red lines) during dehydration of eight uprooted lavender plants under constant temperature (25°C). The number and the letter represent the plant and the acoustic recording system (P: PCI2 and S: Samos), respectively. The dotted lines represent the time at highest peak acoustic activity for two phases ($AE_1$ and $AE_2$).

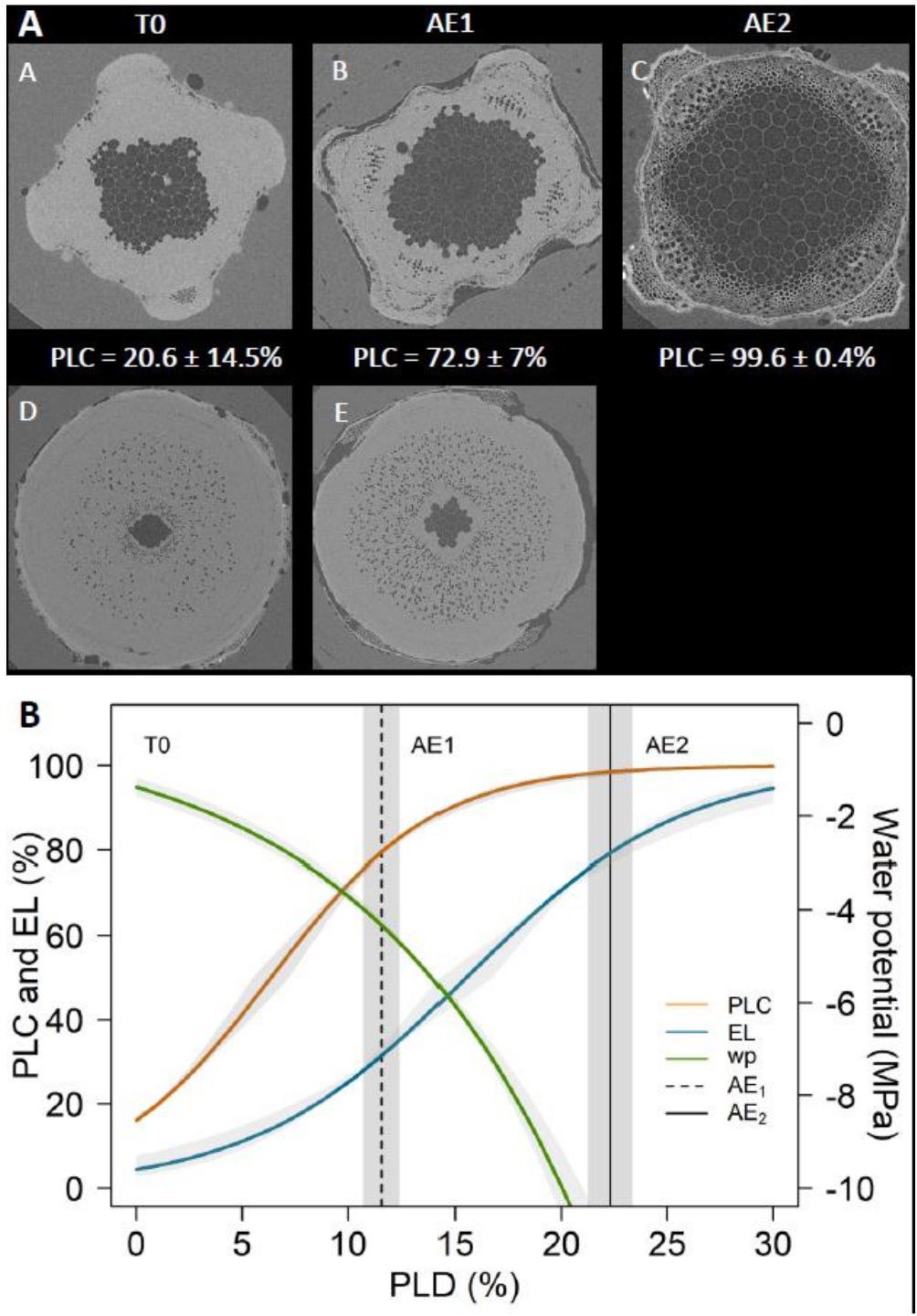

**Figure 2. Upper panel.** Transverse section of inflorescence stems (A-C) and branches (D-E) by high resolution computed tomography at three phases during dehydration: initial phase (T$_0$; Ψ = -1.05 ± 0.12 MPa), after the first peak of acoustic emissions (AE$_1$; Ψ = -4.4 ± 0.03 MPa) and after the second peak of acoustic emissions (AE$_2$; Ψ < -9 MPa). Dark areas represent low-density areas *i.e* embolized vessels and pith. PLC is the loss of xylem hydraulic conductivity for each phase (mean ± SE from n=3 for T$_0$ and AE$_2$, and 5 for AE$_1$). **Lower panel.** PLC, water potential (wp), percent cell lysis (EL) depending on percentage loss of diameter (PLD) during dehydration. PLD at AE$_1$ and AE$_2$ are represented by vertical dotted and black lines, respectively (SE in shaded area). See Lamacque et al. (2020) for further explanations.

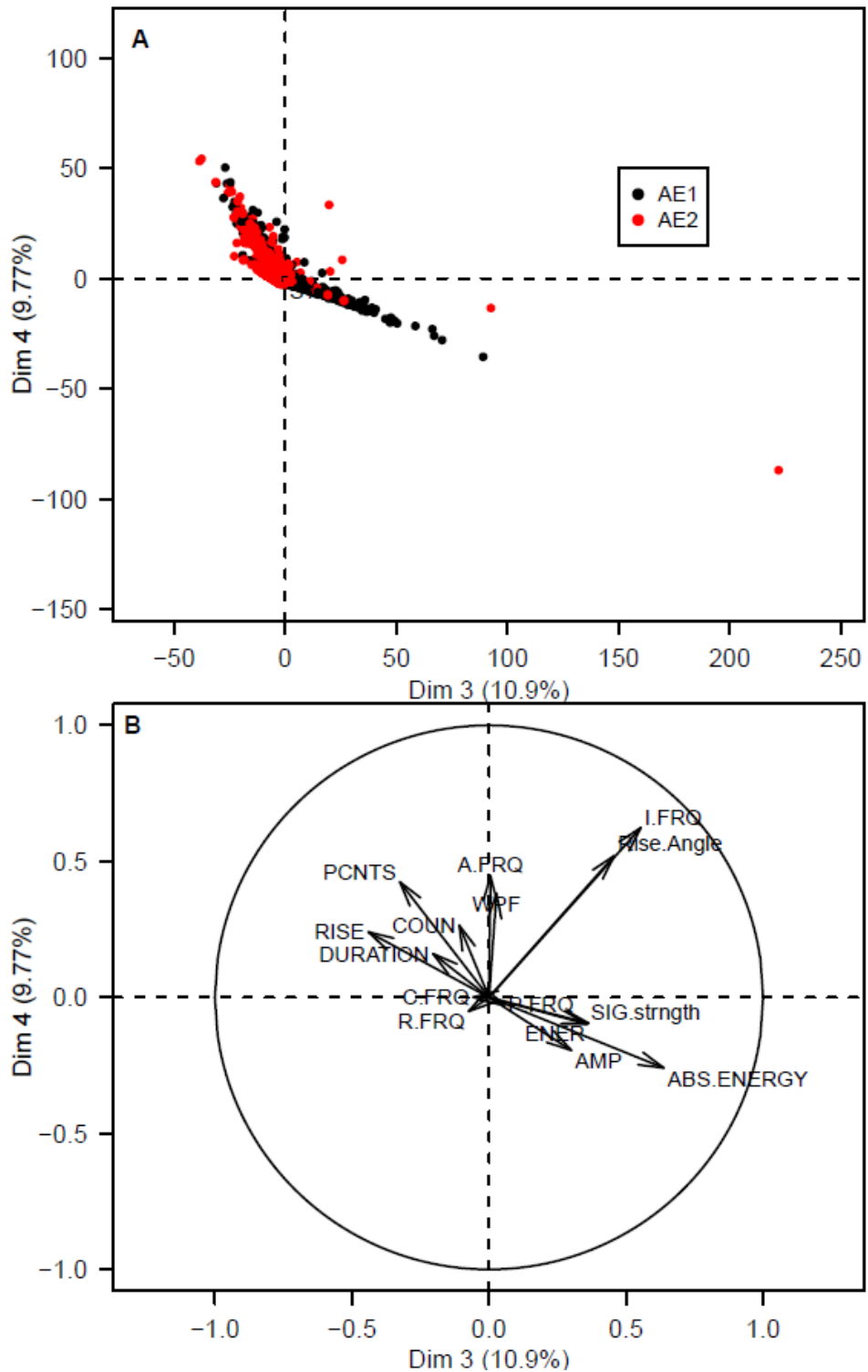

**Figure 3. A.** Principal Component Analysis (PCA) based on characteristics of recorded acoustic emission during dehydration of uprooted lavender with PCI2 system. Black and red dots represent the signals recorded during distinct acoustic phase (AE₁ and AE₂, respectively). Dimensions 3 and 4 are represented as they maximize the discrimination between AE₁ and AE₂ (the other combinations are represented in the supplementary figure S1). **B.** Contribution of each acoustic characteristic to the Dim 3 and 4.

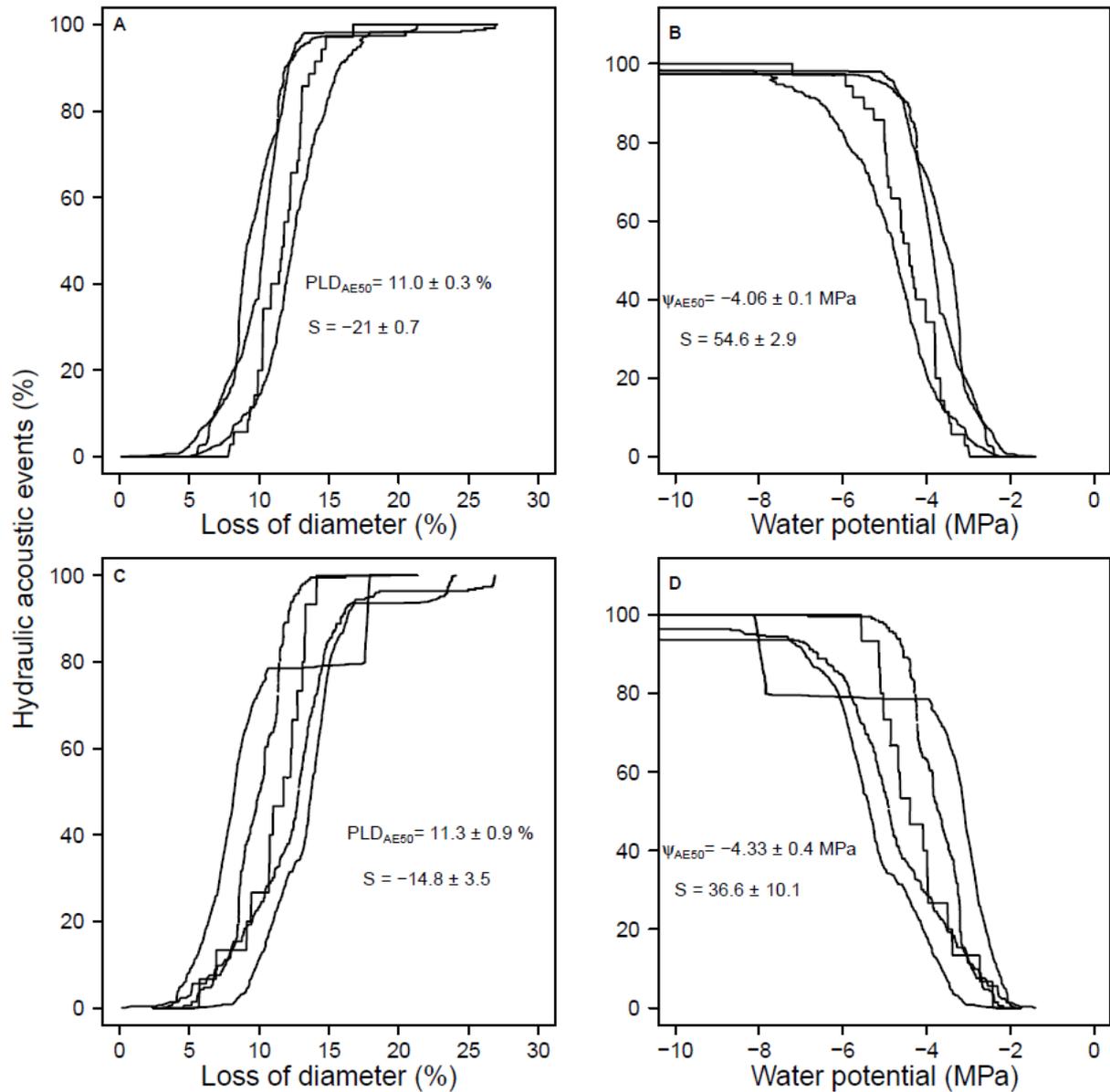

**Figure 4.** Cumulated hydraulically related acoustic events based on Dimension 3 and 4 depending on the percentage loss of diameter (PLD; A and C) and the water potential (B and D). Upper and lower panels represent the signals recorded by PCI2 and Samos recording systems, respectively.

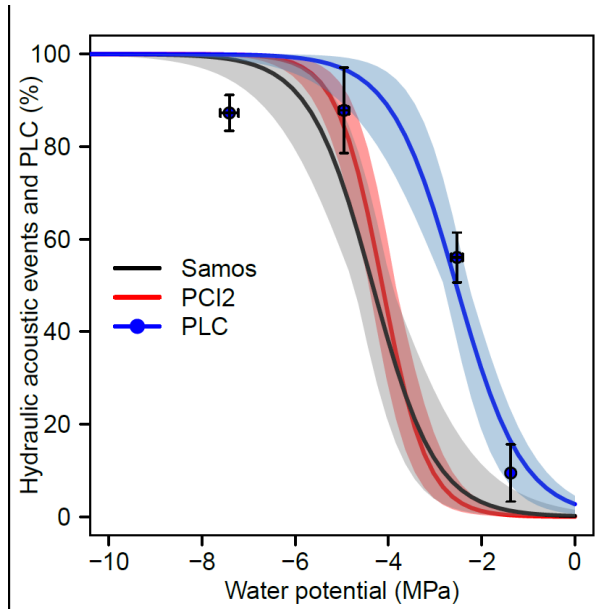

**Figure 5.** Cumulated hydraulically related acoustic events from two independent recording systems PCI2 (black line) and Samos (red line) and loss of hydraulic conductance measured on branches (blue points and line).

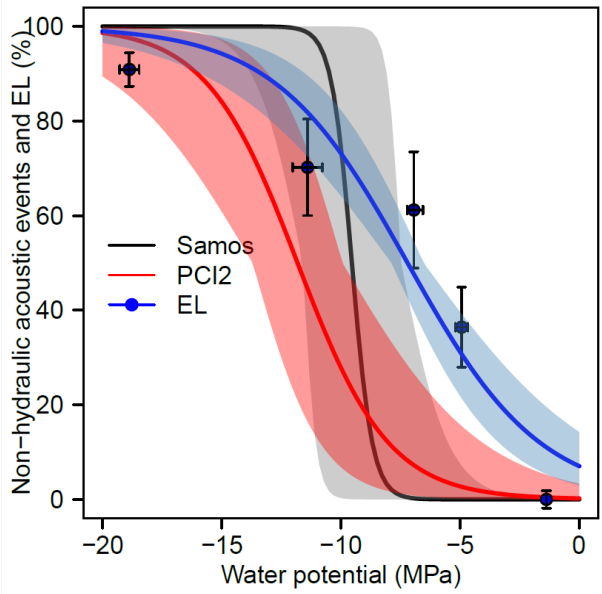

**Figure 6.** Non-hydraulic acoustic events from two independent recording systems PCI2 and Samos, in black and red, respectively. Relative cellular damages measured by electrolyte leakage method during the dehydration are represented in blue.

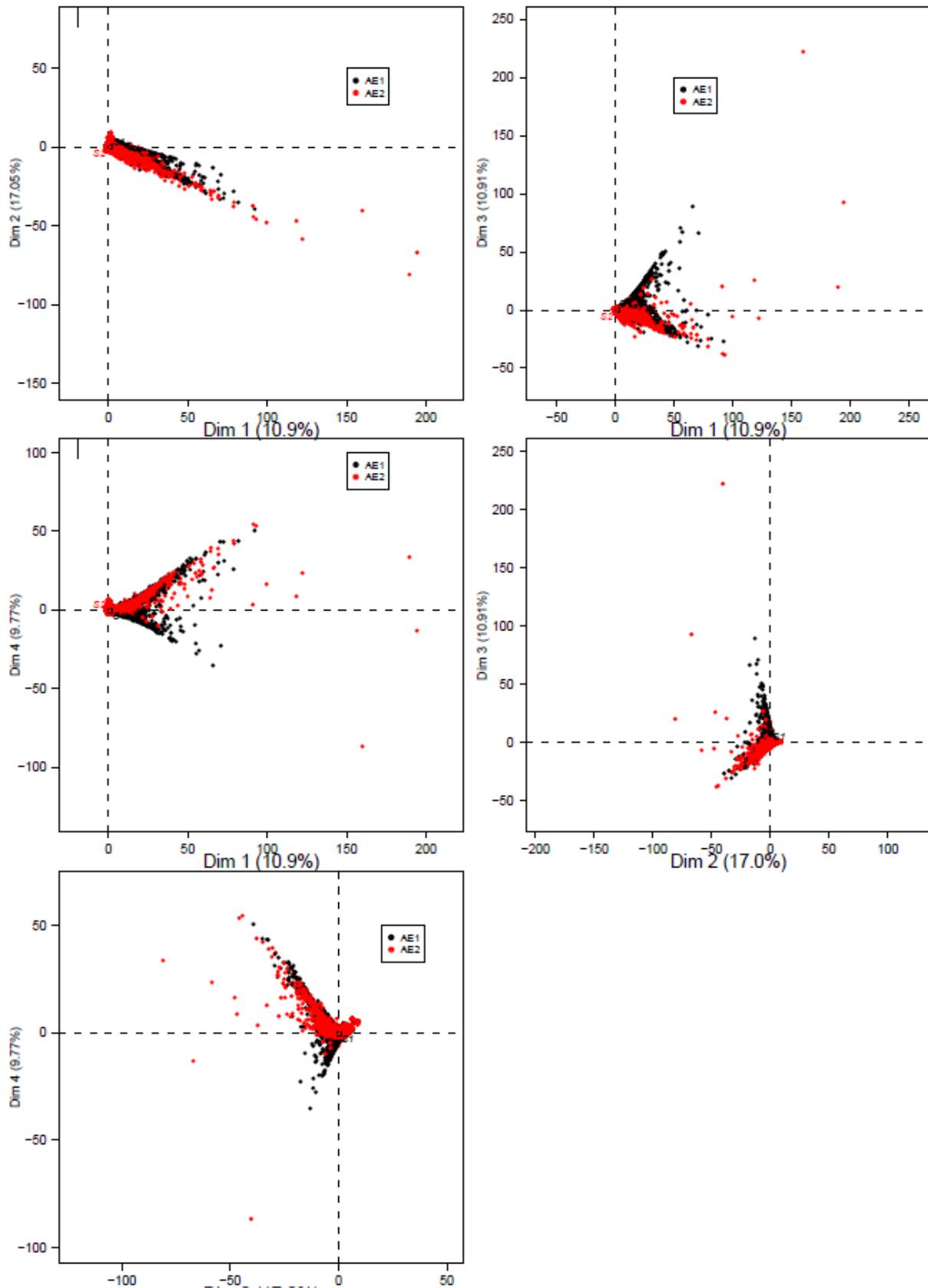

**Supplementary figure S1.** Different combinations of the four first dimensions of the principal component analysis based on characteristics of recorded acoustic emission during dehydration of uprooted lavender with PCI2 system. Black and red dots represent the signals recorded during distinct acoustic phase ($AE_1$ and $AE_2$, respectively).

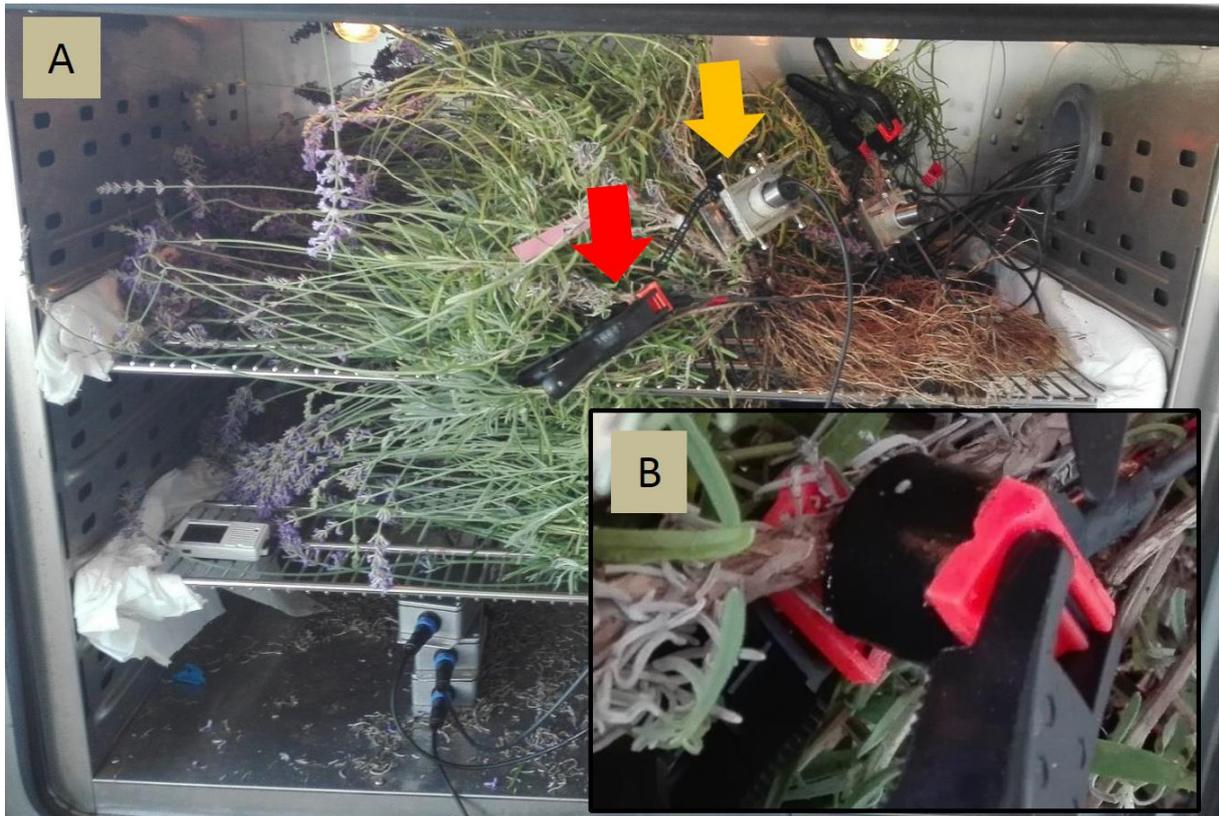

**Supplementary figure S2. A.** Experimental set-up in the temperature-controlled chamber, combining LVDT (orange arrow) and acoustic sensor (red arrow). The inset represent a higher magnification of the acoustic sensor.

**Supplementary table I.** Convolution of the four first latent variables from PCA analysis.

| Latent Variable 1 | | Latent Variable 2 | | Latent Variable 3 | | Latent Variable 4 | |
|---|---|---|---|---|---|---|---|
| Parameter | Contribution | Parameter | Contribution | Parameter | Contribution | Parameter | Contribution |
| **Sig strength** | **0.883** | **WPF** | **0.839** | **Abs Energy** | **0.638** | **I Frq** | **0.624** |
| **Energy** | **0.876** | **A Frq** | **0.759** | **I Frq** | **0.554** | **Rise Time** | **0.521** |
| **Count** | **0.831** | **R Frq** | **0.565** | Rise Time | 0.457 | A Frq | 0.447 |
| **Duration** | **0.784** | P Count | 0.409 | Rise Angle | 0.440 | P Frq | 0.424 |
| **P Count** | **0.644** | Rise Time | 0.401 | Energy | 0.363 | WPF | 0.382 |
| **AMP** | **0.636** | AMP | 0.354 | Sig strength | 0.357 | Count | 0.265 |
| **Rise Time** | **0.593** | Rise Angle | 0.347 | P Frq | 0.325 | Abs Energy | 0.259 |
| Abs Energy | 0.477 | Duration | 0.317 | AMP | 0.300 | Rise Angle | 0.240 |
| P Frq | 0.429 | P Frq | 0.304 | Duration | 0.204 | AMP | 0.194 |
| Rise Angle | 0.415 | C Frq | 0.264 | Count | 0.109 | Duration | 0.160 |
| I Frq | 0.382 | Count | 0.188 | R Frq | 0.074 | Sig strength | 0.098 |
| R Frq | 0.346 | Energy | 0.185 | C Frq | 0.057 | Energy | 0.095 |
| WPF | 0.259 | Sig strength | 0.17 | P Count | 0.055 | R Frq | 0.051 |
| C Frq | 0.204 | I Frq | 0.135 | WPF | 0.027 | P Count | 0.027 |
| A Frq | 0.092 | Abs Energy | 0.078 | A Frq | 0.005 | C Frq | 0.001 |